\documentclass[12pt]{iopart}

\usepackage{graphicx}

\usepackage{upgreek}

\begin{document}
\title{Cooling the Motion of a Silica Microsphere in a Magneto-Gravitational Trap in Ultra-High Vacuum}
\author{Bradley R. Slezak$^1$, Charles W. Lewandowski$^2$, Jen-Feng Hsu$^1$ and Brian D'Urso$^2$}
\address{$^1$ University of Pittsburgh, Pittsburgh PA 15260}
\address{$^2$ Montana State University, Bozeman MT 59715}

\date{\today}

\begin{abstract}
Levitated optomechanical systems, and particularly particles trapped in vacuum, provide unique platforms for studying the mechanical behavior of objects well-isolated from their environment. Ultimately, such systems may enable the study of fundamental questions in quantum mechanics, gravity, and other weak forces. While the optical trapping of nanoparticles has emerged as the prototypical levitated optomechanical system, it is not without problems due to the heating from the high optical intensity required, particularly when combined with a high vacuum environment. Here we investigate a magneto-gravitational trap in ultra-high vacuum.  In contrast to optical trapping, we create an entirely passive trap for diamagnetic particles by utilizing the magnetic field generated by permanent magnets and the gravitational interaction.  We demonstrate cooling the center of mass motion of a trapped silica microsphere from ambient temperature to an effective temperature near or below one milliKelvin in two degrees of freedom by optical feedback damping.

\end{abstract}
\maketitle
\section{Introduction}

Following demonstrations of ground state cooling in clamped mechanical resonator systems~\cite{springboard,teufel,chan}, optical trapping of nanoparticles in vacuum has emerged as a promising technique for pursuing fundamental tests of quantum mechanics~\cite{chang,romeroisart,bateman,thermalsqueeze,vijay}, sensing of weak forces~\cite{gieseler,atto,force,monteiro}, and searches for new physics~\cite{geraci,moore,rider}. However, the high optical intensities required for trapping can result in excessive heating and trapping instabilities at intermediate vacuum~\cite{atto,force,lossmillen,kiesel}. Low total quantum efficiency in detecting the motion has been an obstacle to reaching the quantum ground state. Alternatively, charged microparticles can be trapped with Paul traps or hybrid electro-optical traps~\cite{goldwater,millen,fonseca}, but the required oscillating electric fields come with additonal challenges. 

Here we demonstrate trapping of a silica microsphere in a magneto-gravitational trap in a room-temperature ultra-high vacuum (UHV) environment and cooling the center-of-mass (COM) motion by feedback of the detected movement. Compared to previous results with trapped nanodiamonds in a similar trap~\cite{hsu}, improvements in the trap and the more orientation-independent scattering from microspheres result in at least three orders of magnitude improvement in the effective temperature of the COM motion, reaching temperatures below 1~mK in one degree of freedom. The magneto-gravitational trap utilizes the weak diamagnetism in many common materials, including silica. This trapping mechanism is completely passive; trapped particles can remain confined indefinitely without any external feedback mechanisms. This allows for the use of very low light level illumination, avoiding the heating and instability problems that may occur in optical trapping experiments as residual gas pressure is reduced~\cite{atto,force,lossmillen,kiesel}.

A hybrid magneto-gravitational trap was designed to utilize magnetic and gravitational forces to trap diamagnetic particles in a weak three-dimensional potential well.  The potential energy of a particle in the presence of an external magnetic field and the earth's gravitational field is given by: 
\begin{equation}
\label{eq:U}
U= -\frac{\chi B^2 V}{2\mu_0} + mgy, 
\end{equation}
where $\chi$, $V$, and $m$ are the magnetic susceptibility, volume, and mass of the particle, respectively; $g$ the acceleration due to gravity, $y$ the vertical position of the particle, and $\mu_0$ the vacuum permeability.  For diamagnetic particles ($\chi<0$), the minimum in the potential energy due to the magnetic field occurs at a magnetic field minimum, and the particle can be trapped. To construct a trap, we start by considering a linear magnetic quadrupole field, centered on and invariant under translations along the $z$ axis.  Such a field confines the particle in the transverse ($x$) and vertical ($y$) directions, by virtue of a magnetic field zero along the line $x = y = 0$, but the particle is unconstrained in the axial ($z$) direction.  To make a full three-dimensional trap, we distort the geometry so the region of zero field is curved upward in the $y$-$z$ plane as $|z|$ increases. The motion of the particle is still constrained by the magnetic field to remain near this zero-field region in the $x$ and $y$ directions, and the earth's gravity makes the potential energy minimum occur at $z=0$, creating a full three-dimensional trap.

\begin{figure*}[ht!]
\includegraphics{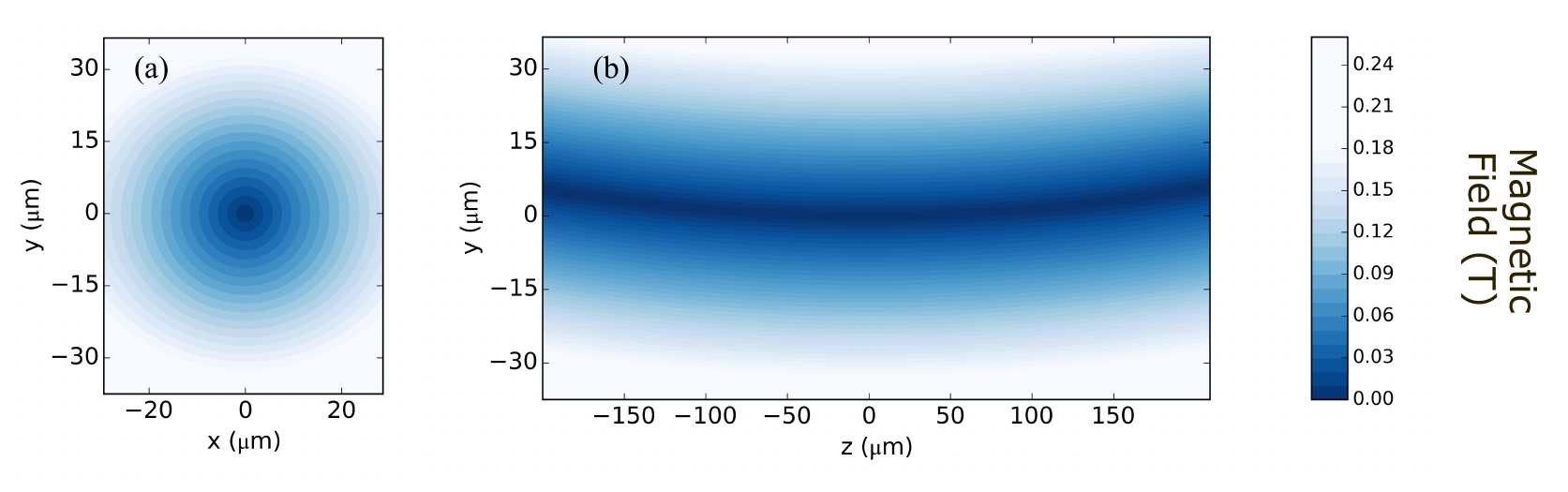}
\caption{\label{fig:b_field} Plots of the modeled magnetic field magnitude based on a spherical harmonic expansion with coefficients chosen to match the observed trap frequency. (a) Magnetic field strength for slice of axial plane ($z=0$).  Quadrapole nature is evident as $\left| B \right| \propto r$ at the center. (b) Magnetic field strength for slice at transverse plane ($x=0$). Upward curvature of the zero field region results in the trapping force in the $z$ direction which prevents the particle from escaping the trap.}
\end{figure*}

To model this field for analysis, we expand the magnetic scalar potential in spherical harmonics. We consider only the first three non-zero multipole moments, based upon the trapping region symmetries, with relative magnitudes determined by experimental data of trap stiffness in each direction~\cite{hsu}. After selecting the coefficients in the expansion to match the observed particle COM oscillation frequencies, we plot the trapping field.  Fig.~\ref{fig:b_field}(a) shows a slice of the model magnetic field magnitude in a plane where the confinement is purely magnetic, while Fig.~\ref{fig:b_field}(b) shows the subtle change in height of the zero field region as the axial distance from the center increases.

\begin{figure}[ht!]
\includegraphics{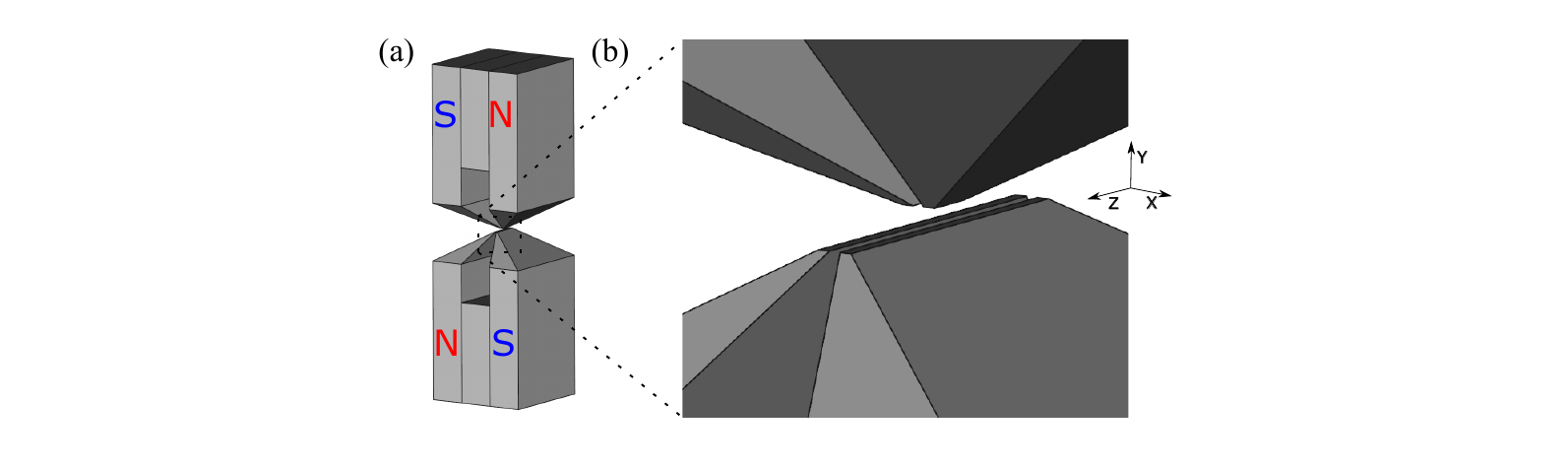}
\caption{\label{fig:cam} (a) Drawing of the the magneto-gravitation trap construction.  SmCo permanent magnets are sandwiched between top and bottom pairs of ferromagnetic pole pieces; the pole pieces are arranged in a quadrupole configuration. (b) Close up of trapping region, showing the defined axes (tranverse ($x$), vertical ($y$), axial ($z$)), and narrower cut of the top pole pieces to break quadrupole symmetry and curve the zero-field region upward for axial trapping.}
\end{figure}

The implemented trap uses four pole pieces in a quadrupole-like configuration, machined from a ferromagnetic material with a high saturation magnetization (Hiperco 50A, Ed Fagan Inc.). Two SmCo permanent magnets are placed between the pole piece pairs to generate the magnetic field (see Fig.~\ref{fig:cam}).  To create the upward curvature shown in Fig.~\ref{fig:b_field}(b), the top pole pieces are cut shorter in the axial direction than the bottom pole pieces, breaking the quadrupole symmetry, as evident in Fig.~\ref{fig:cam}(b).  Magnetic field gradients near the center of the trapping region are $\sim 10^4$~T/m.

The equilibrium position of the particle is at $x=z=0$ by symmetry, but biased to $y<0$ due to earth's gravity. The total trapping potential can be expanded around the equilibrium position in a power series including anharmonic terms; by considering only small displacements compared to the size of the trap, we treat the potential as purely harmonic, with independent oscillation frequencies in each direction. Thus, we consider the classical harmonic oscillator Hamiltonian:
\begin{equation}
H = \sum_i\frac{1}{2}mv_i^2 + \frac{1}{2}m\omega_i^2x_i^2,
\label{eq:prob} 
\end{equation}
where $v_i$ is the velocity of the particle, $\omega_i$ the natural angular frequency, $x_i$ is the displacement from equilibrium, and the subscript $i=x, y, z$ denotes the degree of freedom. For the trap configuration discussed in this paper, $\omega_x = 2\pi \times  59.6$~Hz, $\omega_y = 2\pi \times 96.9$~Hz, and $\omega_z = 2\pi \times 7.01$~Hz. We utilize this harmonic approximation for the remainder of the analysis.  

The data presented in this work originate from a single microsphere; analyses for other single microspheres provide similar results.

\section{Methods}

The particles utilized are silica (SiO$_2$) microspheres (Cospheric LLC SiO2MS-1.8~1.54um-1g), with a mean diameter of 1.54~$\upmu$m and \%CV=9.6 (with \%CV defined as the standard deviation of the mean divided by the mean, expressed as a percentage).  Particles are loaded into the trap by ultrasonically shaking them off the tip of a horn \cite{horn} and into the trap at atmospheric pressure.  By scattering light off the particles and imaging them with a 0.50~NA objective, we are able to distinguish between single particles and clusters of particles of this size, and proceed only with single particles for experiments.

Typically, several single particles or clusters are trapped in any load attempt; electrostatic repulsion due to incidental charge on the particles largely prevents them from agglomerating. Since each particle could have a different charge to mass ratio, we can reduce the number of particles in the trap by placing a small (0.5-2~V), slowly oscillating (0.5-1~Hz) voltage between the top and bottom pole pieces; this drives particles with high charge to mass ratios out of the trap.  Additionally, this process allows us to identify the sign of a particle's charge by monitoring its phase response to the drive.  The driving is repeated until a single sphere with the desired charge polarity is left in the trap, or the loading process is repeated if necessary.  Once loaded, microspheres remain trapped indefinitely without any external feedback mechanisms from atmospheric pressure to UHV.

To neutralize the microsphere, an ionizing radiation source (Americium-241) is brought near a selectively trapped positively charged particle at atmospheric pressure.  We monitor the response of the electrically driven motion of the microsphere on a camera while the source is nearby until no motion is evident with the voltage increased up to $\sim 10$~V. At this point, there may still be a small amount of residual charge on the particle.  

\begin{figure}[h]
\includegraphics{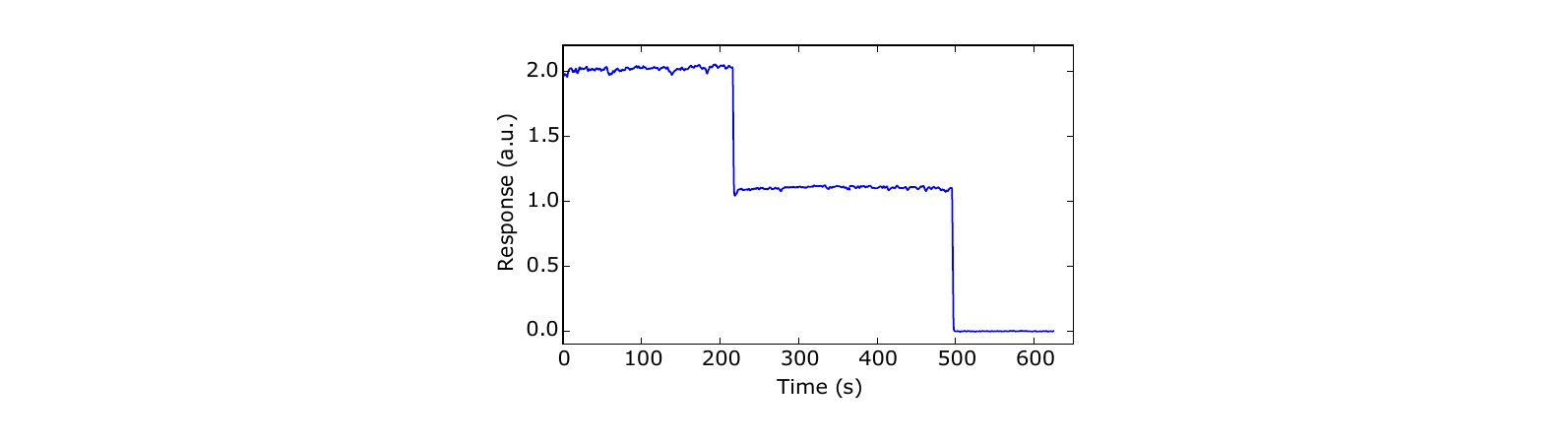}
\caption{The magnitude of the response of the particle to an oscillating electric field as the trapping surface is exposed to UV light.  Each step in the response is the result of a single electron being added to the initially positively charged microsphere.}
\label{fig:charge}
\end{figure}

To completely neutralize the particle, we expose the nearby pole pieces to ultra-violet (UV) light. The light includes photons with sufficient energy to knock electrons off of the nearby pole pieces and into the particle.  In high vacuum, we monitor the amplitude of the electrically driven motion of the microsphere on a lock-in amplifier and observe discrete steps in the output, corresponding to single electron additions to the particle, which is similar to the method presented in~\cite{charge}.  The response of the particle during this process is shown in Fig.~\ref{fig:charge}.  The rate of charging is sufficiently slow that we can extinguish the light while the response amplitude is zero.

\subsection{Mass Calibration and Ultra-High Vacuum}
After the particle is sufficiently discharged, we slowly pump the chamber from atmospheric pressure to $\sim 10^{-2}$~Torr, followed by faster pumping to $\sim 10^{-4}$~Torr by briefly opening the chamber to a turbomolecular pump.  At this moderate vacuum, the particles have a fast ($\sim 10$~s) thermalization time.  Once the particle is in thermal equilibrium with its environment, the total energy of the particle follows a Boltzmann distribution of energies, $P(E) \propto \exp(-E/k_BT)$, where $k_B$ is Boltzmann's constant and $T$ is the temperature of the environment.   Substituting the classical harmonic oscillator Hamiltonian for the energy $E$, and limiting the analysis to a single degree of freedom only, we write the probability distribution for the particle's position:
\begin{equation}
P(x_i)\propto \exp\left(-\frac{m\omega_i^2x_i^2}{2k_BT}\right)
\label{eq:proby}
\end{equation}
To measure $\omega_i$, we Fourier transform position data obtained from imaging the thermally-driven particle onto a quadrant photodiode detector (discussed in further detail in Section~\ref{sec:detection}) and obtain a fit for the center frequency.  

We monitor the thermal motion of the particle using a sequence of images from a high speed camera, which provides a sampling of the probability distribution in Eq.~\ref{eq:proby}. From this data, we extract the mass of the particle by fitting the distribution with $m$ as a fit parameter.  For this process, we utilize both the vertical ($y$) and axial ($z$) degrees of freedom, and then average the extracted masses.  The mass of the particular microsphere extracted for the data presented is $m=(3.10 \pm  0.05) \times10^{-15}$~kg, which is well within the range of expected masses, $m = (3.4 \pm 1.0) \times 10^{-15}$~kg, from the provided distribution on particle diameter. This provides verification that we have indeed trapped a single particle.

\begin{figure}[h]
\includegraphics{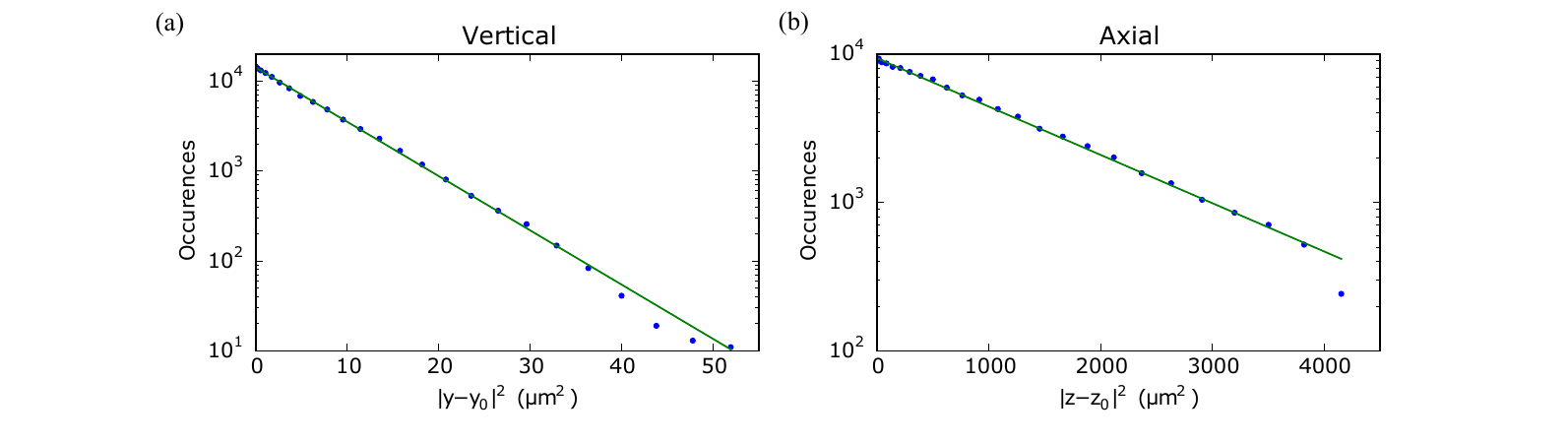}
\caption{Histograms of squared displacement of a single trapped particle in the (a) vertical ($y$) and (b) axial ($z$) degrees of freedom, with COM motion thermalized with the ambient temperature.  The mass of the trapped particle is extracted from each distribution and averaged for the reported mass.}
\label{fig:hist}
\end{figure}

We then pump the trap chamber to high vacuum by opening the chamber to a turbomolecular pump. The chamber and trap are then baked (custom PEEK gaskets \cite{peek}, suitable for baking, were utilized for glass to metal vacuum seals), with the particle remaining trapped without observation or correction. For this work, the chamber was baked to 135$^{\circ}$C for 24 hours.  We then open the chamber to an ion pump and disconnect the turbomolecular pump to eliminate vibrations. The final chamber pressure achieved is $\sim 2.0 \times 10^{-10}$~Torr at ambient temperature. No special precautions are needed to keep the particle trapped and stable from the initial pumpdown to UHV, and the particle remains trapped indefinitely under UHV.

\subsection{Detection and Optical Feedback Cooling}\label{sec:detection}
\begin{figure}[ht!]
\includegraphics{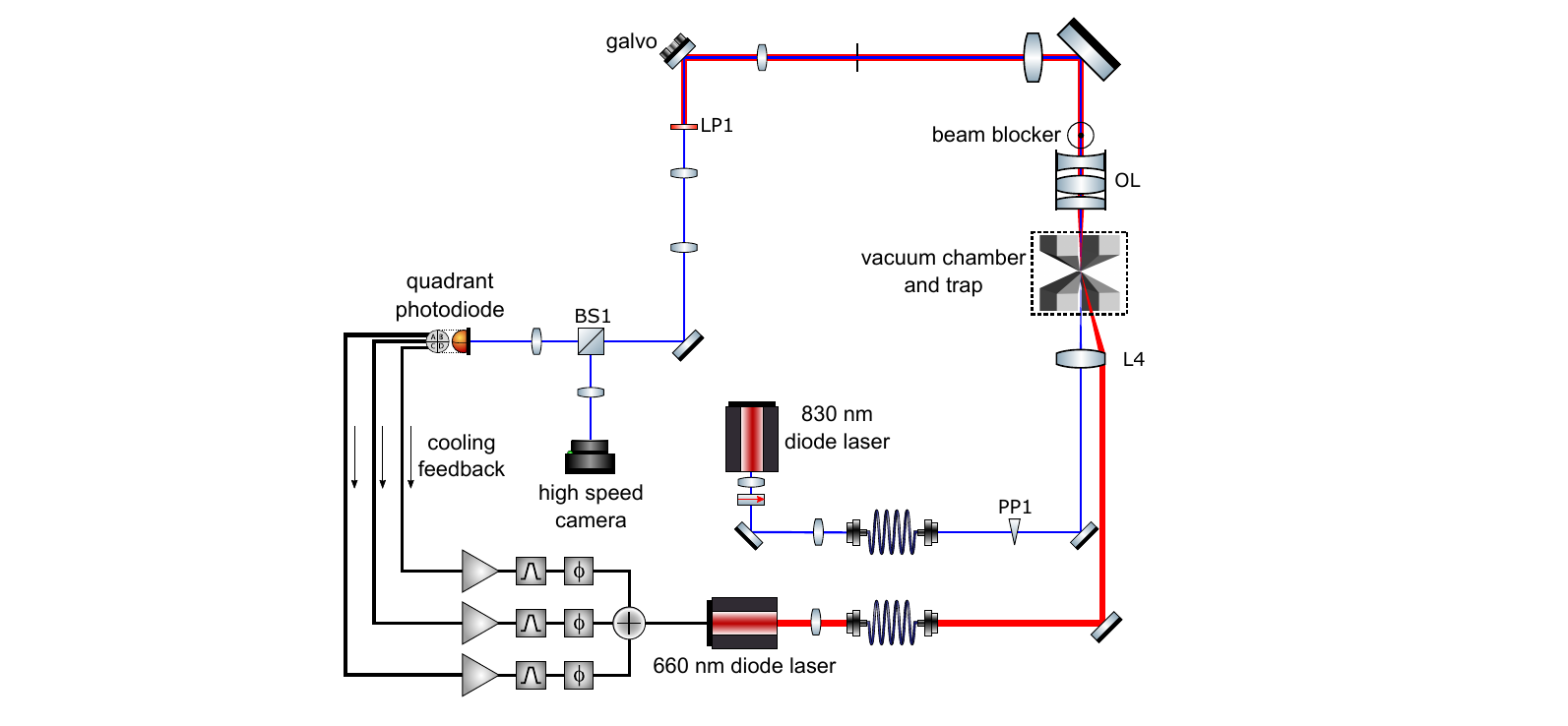}
\caption{\label{fig:Optics}  Illumination and detection path via the 830~nm diode laser is portrayed by the blue path, with transmitted illumination blocked at the beam blocker.  Light is relayed through two pairs of lenses and then split onto the high speed camera and quadrant photodiode. Light from a 660~nm control laser, used for driving the particle via radiation pressure, is portrayed by the red path and blocked by long pass filter LP1.}
\end{figure}

Illumination is accomplished by loosely focusing 830~nm light from a diode laser into the trap from the transverse ($x$) direction.  Both light transmitted through the trapping region and light scattered from the particle are collected via a home-made 0.50~NA objective (lens triplet OL in Fig.~\ref{fig:Optics}).  

\begin{figure}[ht!]
\includegraphics{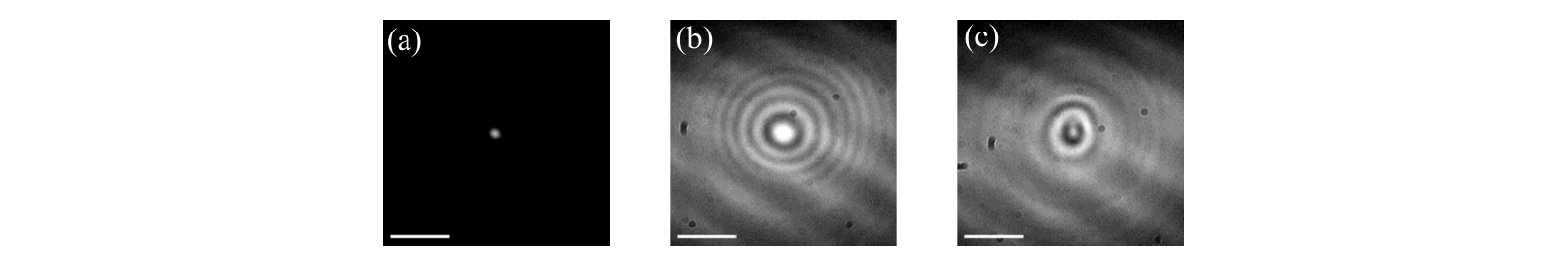}
\caption{\label{fig:pics} (a) Dark field image of a silica microsphere obtained by blocking the illumination source at the back plane of the objective (b) Bright field image obtained by imaging the illumination light with the scattered light.  Interference between the transmitted and scattered light causes a bright particle center on one side of the objective's focus (c) Bright field image similar to (b), but with the objective defocused to the opposite side, showing a change to a dark particle center.  The scale bars are 10~$\upmu$m.}
\end{figure}

This illumination scheme allows for a dark-field image of the particle, as shown in Fig.~\ref{fig:pics}(a), by blocking the transmitted illumination light at the back of the objective lens and imaging the light onto a high speed camera or quadrant photodiode.  The transmitted light could also be retained, allowing both scattered and transmitted light to create a brightfield image, as shown in Fig.~\ref{fig:pics}(b) and Fig.~\ref{fig:pics}(c).  These brightfield images contain interference fringes of the scattered light with the transmitted light, which are visible as a dark or bright center surrounded by interference rings.  This contains information on the location of the particle in the transverse ($x$) direction as it moves in and out of the focus of the objective, and could be utilized as a detection method for cooling this degree of freedom.

Three electrical signals are generated by (dark-field) imaging the microsphere onto a quadrant photodiode;  left-right and top-bottom difference currents yield signals proportional to the displacement of the particle in the $y$ and $z$ directions, respectively. A third signal summing all quadrants can be used to sense motion along the optical axis ($x$). The electrical signals are amplified, then phase-shifted and band-pass filtered around the resonance frequencies by a microcontroller in real time. The phase shift is chosen so that they are proportional to the velocity of the microsphere in each direction, and the resulting signals are used to modulate the amplitude of a second diode laser at 660~nm. The amplitude modulated 660~nm light is loosely focused on the microsphere, providing a damping (or driving) force via radiation pressure.  This beam is translated off the center of its focusing lens L4, resulting in non-zero components of the force in all directions. All COM degrees of freedom can be cooled with this single laser, since the frequency of each degree of freedom is well separated and only sensitive to a force oscillating near its own resonance frequency.

\section{Results}
We report significant cooling of the mechanical motion in both the vertical ($y$) and axial ($z$) directions by feedback of the detected motion. In order to assign an effective temperature $T_i^\prime$ for either degree of freedom, we utilize the simple relationship:
\begin{equation}
\frac{1}{2}k_B T_i^\prime = \frac{1}{2}m\omega_i^2\langle x_i^2\rangle
\label{eq:teff}
\end{equation}
Since we computed $m$ from the thermalized motion, we can easily measure $\omega_i$ and $\langle x_i^2 \rangle$ from the voltage output of the quadrant photodiode circuit, which provides the particle's motion as a function of time.  The voltage is calibrated to a distance measurement by utilizing the calibrated magnification of images on the high speed camera as a reference.

\begin{figure}[ht!]
\includegraphics{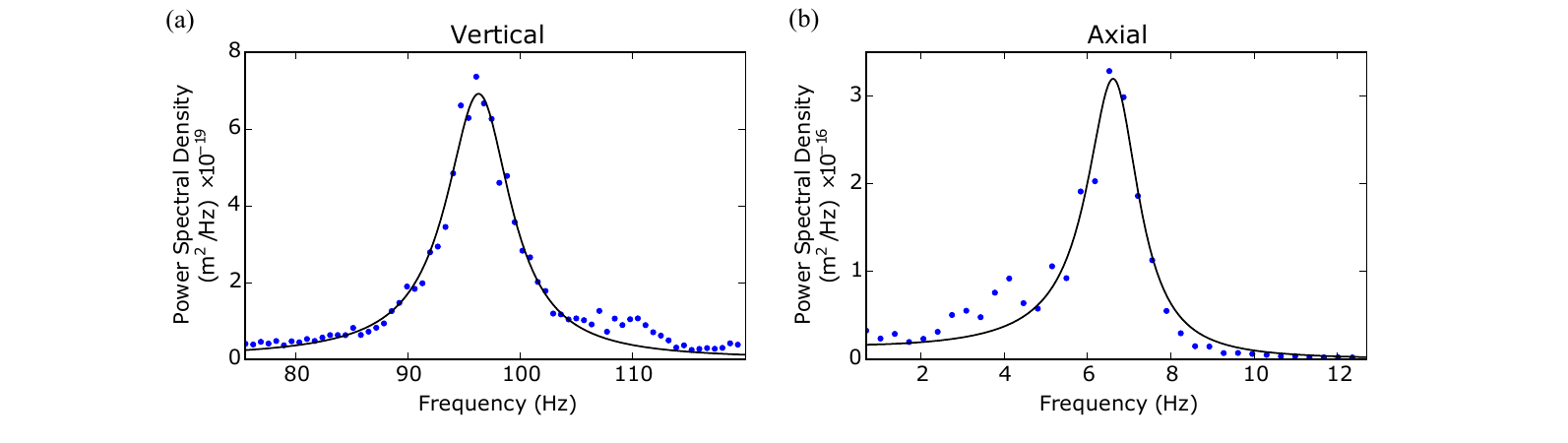}
\caption{\label{fig:spec} (a) Plot of the power spectral density of the vertical motion near the cooled resonance frequency $\omega_y = 2\pi\times 96.5$~Hz. (b) Plot of the power spectral density of the axial motion near the cooled resonance frequency $\omega_z = 2\pi\times 6.7$~Hz. Both include a fit to the expected response of a classical, damped, noise-driven oscillator.}
\end{figure}

We calculate the effective temperature by fitting the power spectrum of the motion to the theoretical power spectral density function for a damped harmonic oscillator:
\begin{equation}
\mathit{PSD}(A,f_i,\gamma) = A \left((f_i^2-f^2)^2 + \gamma^2f^2\right)^{-1}
\label{eq:psd}
\end{equation}
with $\gamma$ the width of the resonance, $f_i=\omega_i/2\pi$, and A an overall scaling factor.
  As a check, we also directly calculate $\langle x_i^2 \rangle$ from data filtered around the resonance frequency.  The fits for the axial and vertical data can be seen in Fig.~\ref{fig:spec}.  The fits yield an effective temperature in the vertical direction of $T_y^\prime = 1.2 \pm 0.1$~mK, and an effective temperature in the axial direction of $T_z^\prime = 0.6 \pm 0.1$~mK.  The direct calculation of $\langle x_i^2 \rangle$ gives similar results.  Typically we cool the transverse motion weakly at the same time, but do not calibrate the response.

\section{Discussion}
The cooling results for the vertical motion ($T_y^\prime=1.2$~mK) and axial motion ($T_z^\prime=0.6$~mK) are similar to the coldest temperatures reported for the motion of a particle in an optical trap~\cite{vijay,milligieseler,li}. However, since the oscillation frequencies in the magneto-gravitational trap are much lower than the typical frequencies in optical traps, the motion in the magneto-gravitational trap is much further from the quantum ground state. In particular, we can calculate the average number state for each cooled degree of freedom $\bar{n}_i^\prime$ as 
\begin{equation}
\bar{n}_i^\prime=\frac{k_B T_i^\prime}{\hbar \omega_i},
\end{equation}
which gives $\bar{n}_y^\prime=2.5\times10^5$ and $\bar{n}_z^\prime=1.9\times10^6$, indicating the motion is still far from the quantum ground state.

The measured temperatures of the vertical and axial motions ($T_y^\prime=1.2$~mK, $T_z^\prime=0.6$~mK), combined with their cooled linewidths ($\Gamma_y^\prime/2\pi = 7.0$~Hz, $\Gamma_z^\prime/2\pi = 1.5$~Hz), can be used to place a limit on the natural damping rate of the motions in the trap, $\Gamma_i$. For each degree of freedom, we have
\begin{equation}
\frac{T_i^\prime}{T} \geq \frac{\Gamma_i}{\Gamma_i^\prime},
\end{equation} 
where the environment temperature $T=295$~K and equality is expected for a classical system with no additional heating mechanism and no noise in the detector or feedback loop. This can be solved to yield limits on the natural damping rates $\Gamma_y/2\pi \leq 3 \times 10^{-5}$~Hz and $\Gamma_z/2\pi \leq 3 \times 10^{-6}$~Hz. An estimate for the natural damping rate based on the residual gas pressure \cite{collisions} is determined as $\Gamma/2\pi = 2 \times 10^{-8}$~Hz, suggesting that vacuum is not currently the primary limitation. We have not yet determined the origin of the excess heating and/or damping.

\section{Conclusion}
We have demonstrated cooling of two degrees of freedom of the mechanical motion of a silica microsphere in a magneto-gravitational trap from room temperature to near or below 1~mK, with precise determination of the trapped particle mass. While the motion is not yet approaching the quantum ground state, this system has several unique features that make it attractive for continued development, including passive stability, compatibility with UHV, and extremely low natural damping rates. Further improvements in the vacuum, isolation from environmental disturbances, and increased feedback cooling strength may provide a path to perform experiments in the quantum mechanical regime in these traps.

\section{References}

\nocite{*}

\bibliographystyle{unsrt}
\bibliography{SingleCoolBib}
\end{document}